\documentclass[pre,aps ,superscriptaddress,showpacs]{revtex4-1}
\usepackage{natbib}
\usepackage{amsmath}
\usepackage{epsfig}
\usepackage{color}

\DeclareMathOperator*{\sumint}{%
\mathchoice%
{\ooalign{$\displaystyle\sum$\cr\hidewidth$\displaystyle\int$\hidewidth\cr}}
{\ooalign{\raisebox{.14\height}{\scalebox{.7}{$\textstyle\sum$}}\cr\hidewidth$\textstyle\int$\hidewidth\cr}}
{\ooalign{\raisebox{.2\height}{\scalebox{.6}{$\scriptstyle\sum $}}\cr$\scriptstyle\int$\cr}}
{\ooalign{\raisebox{.2\height}{\scalebox{.6}{$\scriptstyle\sum$}}\cr$\scriptstyle\int$\cr}}
}

\begin{document}

\title{Stochastic Floquet quantum heat engines and stochastic efficiencies}
\author{Fei Liu}
\email[Email address: ]{feiliu@buaa.edu.cn}
\affiliation{School of Physics, Beihang University, Beijing 100191, China}
\author{Shanhe Su}
%\email{Email address: shanheSu@xiameng.edu.cn}
\affiliation{Department of Physics, Xiamen University, Xiamen 361005, China}

\date{\today}

\begin{abstract}
{Based on the notion of quantum trajectory, we present a stochastic theoretical framework for Floquet quantum heat engines. As an application, the large deviation functions of two types of stochastic efficiencies for a two-level Floquet quantum heat engine are investigated. We find that the statistics of one efficiency agree well with the predictions of the universal theory of efficiency fluctuations developed by Verley et al. [Phys. Rev. E {\bf 90}, 052145 (2014)], whereas the statistics of the other efficiency do not. The reason for this discrepancy is attributed to the lack of fluctuation theorems for the latter type of efficiency.    }

\end{abstract}
\maketitle

\section{Introduction}
\label{section1}

Since Scovil and Schulz-Dubois~\cite{Scovil1959} first described a quantum heat engine 60 years ago, the notion of this engine has attracted considerable interest~\cite{Kosloff2013,Kosloff2014,Alicki2018}. With great achievements in microfabrication and microcontrolling, the idea of the quantum heat engine is gradually becoming reality. Recently, a single-atom heat engine~\cite{Rossnagel2016} and a quantum heat engine with energy fluctuations~\cite{Peterson2019} have been successfully implemented.%Editor: Please ensure that the intended meaning has been maintained in this edit to 'successfully'.

A large number of models of quantum heat engines have been proposed over the past couple of decades, and various aspects of quantum heat engines have been investigated, e.g., the quantum supremacy of quantum working media~\cite{Zhang2007,Jaramillo2016,Campisi2016,Ma2017,Klatzow2019} and surpassing the Carnot efficiency using nonequilibrium heat baths~\cite{Scully2003,Huang2012,Robnagel2014,Abah2014,Alicki2015,Manzano2016,Klaers2017,
Agarwalla2017,Ghosh2018}. According to their operating modes, these models are roughly classified into reciprocating engines and continuous engines. The former engines are typically composed of four processes and operate analogously to the classic Carnot cycle or Otto cycle~\cite{Sisman1999,Feldmann2000,He2002,Quan2007,Zhang2007,Segal2008,Huang2012,Wang2013,Allahverdyan2013,
Abah2014,Robnagel2014,Jaramillo2016,Manzano2016,Campisi2016,Ma2017}, while the latter engines operate by permanently contacting multiple heat baths with different temperatures and continuously converting the heat absorbed from those baths into work~\cite{Alicki1979,Kosloff1984,Geva1992,Levy2012,Gelbwaser2013,Szczygielski2013,GelbwaserKlimovsky2015,
Alicki2015,Agarwalla2017}. The dynamics of continuous heat engines are usually modeled by Markov quantum master equations~\cite{Gorini1976,Davies1974,Lindblad1975,Spohn1978,Breuer2002,Alicki2010,Rivas2012}.

The aim of this paper is to present a stochastic theory for continuous quantum heat engines that are specially described by the Floquet quantum master equation~\cite{Bluemel1991,Kohler1997,Breuer1997,Alicki2006}. This equation was originally developed for periodically driven open quantum systems and has recently been applied to quantum heat engines~\cite{Kosloff2013,Szczygielski2013,Gelbwaser2013,GelbwaserKlimovsky2015,Alicki2015}. The Floquet quantum master equation has two obvious advantages. On the one hand, the periodically driven character of this equation is highly consistent with the cyclic operation of continuous heat engines. On the other hand, this equation can be rigorously derived by open quantum system theory~\cite{Breuer2002,Alicki2010,Rivas2012}. However, previous studies have mainly focused on the average performances of quantum heat engines. Considering that quantum and thermal fluctuations are intrinsic and are not negligible at the nanoscale~\cite{Rossnagel2016,Klaers2017,Peterson2019,Klatzow2019}, constructing stochastic theories for quantum heat engines is essential. Cuetara et al.~\cite{Cuetara2015} have already taken a step in this direction. Based on stochastic heat defined by the two-energy-measurement scheme~\cite{Kurchan2000,Talkner2007,Campisi2011} performed on heat baths, Cuetara et al. derived a modified quantum master equation~\cite{Esposito2009,Gasparinetti2014} and used it to prove a steady-state fluctuation theorem for the joined mechanical power and heat currents. In addition, they also studied the properties of the stochastic efficiency of a quantum engine that couples two particle reservoirs and converts work into chemical power. The most significant distinction between their stochastic theory and ours is that the notion of quantum trajectory~\cite{Breuer2002,Wiseman2010,Plenio1998} is used herein. The stochastic heat along a quantum trajectory is measured by continuously monitoring the energy changes of the heat baths~\cite{Breuer2003,DeRoeck2006,Derezinski2008,Crooks2008,Horowitz2012,Hekking2013,Liu2014a,Liu2014,Liu2018,Manzano2015,Manzano2018}.
Quantum trajectory provides not only a novel physical perspective on thermodynamic quantities but also some technical advantages. For instance, we may directly obtain statistics by simulating quantum trajectories~\cite{Zoller1987,Molmer93,Breuer2002}. This approach is also more flexible than other methods in exploring new statistical problems.

The remainder of this paper is organized as follows. In Sec.~(\ref{section2}), we introduce the Floquet quantum heat engine and review the Floquet quantum master equation; additionally, the relevant notations are defined therein. In Sec.~(\ref{section3}), a stochastic theory for the Floquet quantum heat engine is presented; in this section, we describe the unravelling of the quantum trajectory, we provide the definitions of stochastic heat and steady-state work along a quantum trajectory, and we introduce two types of stochastic efficiencies and their large deviation functions under long time limits. %Editor: Throughout this paper, please consider replacing 'under long time limits' with 'at long time scales'.
In Sec.~(\ref{section4}), we use a periodically driven quantum two-level system to illustrate the proposed stochastic theory. Section~(\ref{section5}) concludes the paper.

\section{Floquet quantum master equation}
\label{section2}
\begin{figure}
\includegraphics[width=1.\columnwidth]{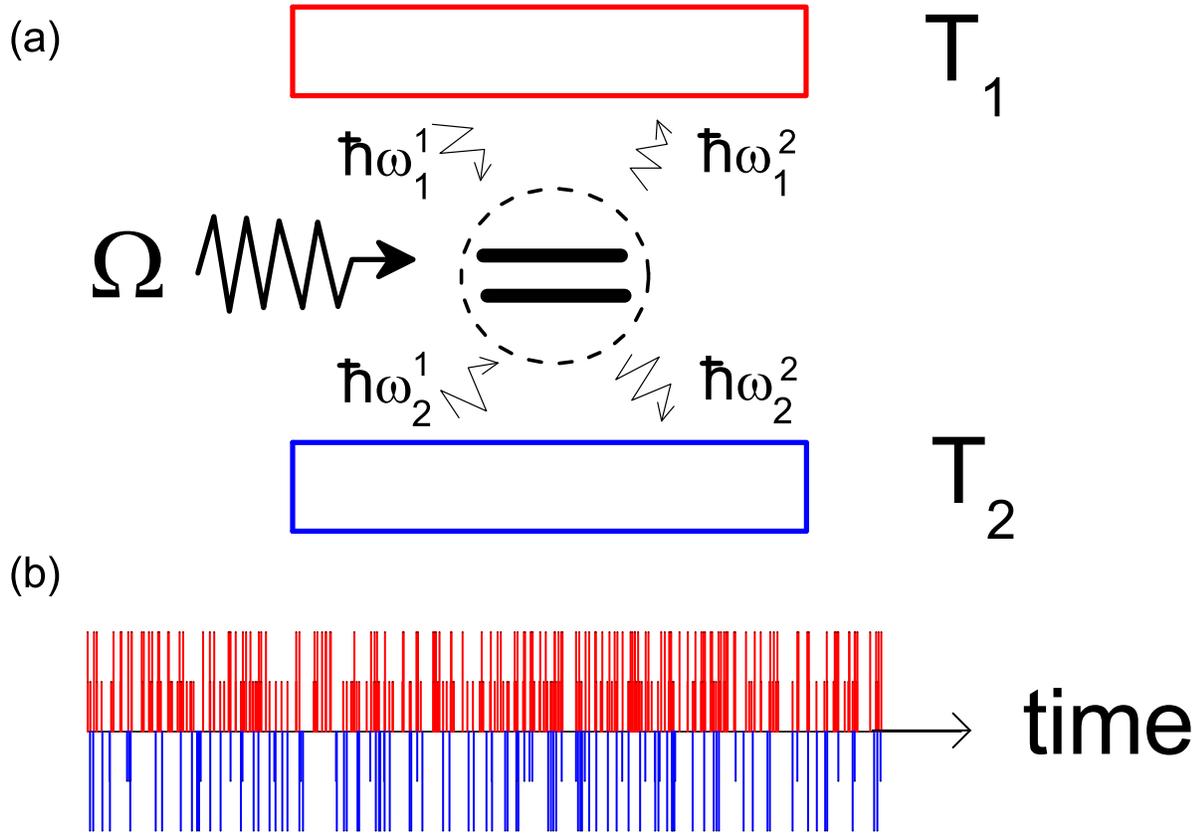}
\caption{(a). Schematic diagram of a Floquet quantum heat engine. The circle is the quantum system driven by a periodic external field with frequency $\Omega$. The small arrowed curves pointing inward and outward the circle denote the absorption and release of quanta or heat, respectively, within the system  from and to the heat baths. (b). A quantum trajectory of the two-level quantum heat engine~(\ref{TLS}) obtained by simulation. The red upper part is the heat exchanged between the system and the hot heat bath with temperature $T_1$, while the blue lower part is the heat exchanged between the system and the cold heat bath with temperature $T_2$. The long and short vertical lines represent the moments of the release and the absorption of heat, respectively. }
\label{fig1}
\end{figure}

Figure~\ref{fig1}(a) shows the schematic diagram of a Floquet quantum heat engine. The quantum system (working medium) couples two heat baths with temperatures $T_1$ and $T_2$, and $T_1>T_2$. The Hamiltonian of the quantum system $H(t)$ is periodically modulated, i.e.,
\begin{eqnarray}
H\left(t+\frac{2\pi}{\Omega}\right)=H(t),
\end{eqnarray}
where $\Omega$ is the driving frequency. Under the assumptions of a weak system-bath coupling condition and time-scale separations~\cite{Breuer1997,Grifoni1998,Alicki2006}, the evolution of the reduced density matrix of the quantum system $\rho(t)$ can be described by the Floquet quantum master equation:
\begin{eqnarray}
\label{FQME}
\partial_{t}\rho(t)=-\frac{i}{\hbar}[H(t),\rho(t)]+D_1(t)[\rho(t)]+D_2(t)[\rho(t)],
\end{eqnarray}
where the $D_k$ term ($k=1,2$) represents dissipation and/or dephasing due to the interaction between the system and the $k$-th bath and is expressed as
\begin{eqnarray}
\label{dissipations}
D_k(t)[\rho]&=&\sum_{\alpha_k=1 }^{N_k}r_k (\omega_k^{\alpha_k})\left[A_k(\omega_k^{\alpha_k},t)\rho A_k^\dag(\omega_k^{\alpha_k},t)- \frac{1}{2}
\left\{A_k^\dag (\omega_k^{\alpha_k},t)A_k(\omega_k^{\alpha_k},t),\rho\right\}\right]. %+ \nonumber \\
%&&\sum_{\omega>0}r_k(-\omega)\left[A_k^\dag(\omega,t)O A_k(\omega,t)- \frac{1}{2}
%\left\{A_k(\omega,t)A_k^\dag(\omega,t),O\right\}\right].
\end{eqnarray}
The summation in the above equation is performed with respect to all possible Bohr frequencies $\omega$, which equal
\begin{eqnarray}
\frac{1}{\hbar}({\epsilon_n-\epsilon_m})+ q\Omega,
\end{eqnarray}
where $q$ are certain integers, and $\epsilon_n$ are the quasi-energy of the Floquet basis $|u_n(t)\rangle$~\cite{Zeldovich1967,Shirley1965}:
\begin{eqnarray}
(H(t)-i\hbar\partial_t )|u_n(t)\rangle =\epsilon_n|u_n(t)\rangle.
\end{eqnarray}
We already think that these quasi-energies are restricted to certain ``Brillouin zone" of width $\hbar\Omega$. Also note that the Floquet basis is periodic with the same frequency $\Omega$. The Bohr frequencies may be positive or negative but always appear in pairs. Because these numbers are determined by the specific interaction operators between the system and the heat baths, we add the subscript $k$ and set the number of Bohr frequencies for the $k$-th bath to $N_k$ and set the positive integers to $\alpha_k=1,\cdots,N_k$.

In Eq.~(\ref{dissipations}), $A_k(\omega_k^{\alpha_k},t)$ and $A_k^\dag(\omega_k^{\alpha_k},t)$ are the Lindblad operators and are related by
\begin{eqnarray}
\label{Hermitiancondition}
A_k(-\omega_k^{\alpha_k},t)=A_k^\dag(\omega_k^{\alpha_k},t).
\end{eqnarray}
Given the system part of the interaction operator between the quantum system and the $k$-th heat bath to be $A_k$, the Lindblad operators are obtained by performing a Fourier-like expansion of the interaction picture operator of $A_k$~\cite{Breuer1997}:
\begin{eqnarray}
U^\dag(t,t_0)A_kU(t,t_0)=\sum_{\alpha_k=1 }^{N_k} A_k(\omega_k^{\alpha_k},t_0)\exp[ -i(t-t_0)\omega_k^{\alpha_k}],
\end{eqnarray}
where $U(t,t_0)$ ($t\ge t_0$) is the time evolution operator of the Hamiltonian $H(t)$ and equal to
\begin{eqnarray}
\sum e^{-i\epsilon_n(t-t_0)/\hbar} |u_n(t)\rangle \langle u_n(t_0)|
\end{eqnarray}
according to the Floquet theory~\cite{Breuer1997}. We can see that Eq.~(\ref{Hermitiancondition}) is a consequence of the Hermitian $A_k$. The last component of the quantum master equation is the assumption that the heat baths are always in their thermal states. Hence, the Fourier transformation $r_k(\omega_k^{\alpha_k})$ of the correlation function of the heat bath component of the interaction operator between the quantum system and the $k$-th heat bath satisfies the Kubo-Martin-Schwinger (KMS) condition~\cite{Breuer2002}:
\begin{eqnarray}
\label{KMScondition}
r_k(-\omega_k^{\alpha_k})=r_k(\omega_k^{\alpha_k})\exp\left(-\beta_k\hbar\omega_k^{\alpha_k}\right),
\end{eqnarray}
where $\beta_k=1/k_BT_k$ and $k_B$ is the Boltzmann constant.
%Our descriptions on the theory are almost the same as previous ones except that here we express the equation in Sch{\"o}dinger picture rather than in the interaction picture~\cite{Szczygielski2013,Kosloff2013} and by using abstract operators~\cite{Breuer1997} rather than in concrete representation~\cite{Cuetara2015}. In Appendix A, we briefly explain the connections between our  and that used by Szczygielski et al.~\cite{Szczygielski2013}.

%These operators can be understood as the eigenoperators of :
%\begin{eqnarray}
%&&[{\cal H}(t),A_k(\omega_k^{\alpha_k},t)]=-\hbar\omega_k^{\alpha_k} A_k(\omega_k^{\alpha_k},t)\\
%&&[{\cal H}(t),A^\dag_k(\omega_k^{\alpha_k},t)]=+\hbar\omega_k^{\alpha_k} A_k^\dag(\omega_k^{\alpha_k},t)
%\end{eqnarray}

\section{Stochastic Floquet quantum heat engine}
\label{section3}
\subsection{Quantum trajectory}
The Floquet quantum master equation~(\ref{FQME}) can be unravelled into quantum trajectories. Considering that this stochastic theory is a straightforward extension of the theory of a single heat bath~\cite{Liu2016a,Liu2018}, here, we present only a brief explanation. First, Eq.~(\ref{FQME}) is rewritten as
\begin{eqnarray}
\label{peturbedmastereq}
\partial_t \rho=L_0(t)[ \rho ] +\sum_{k=1}^2\sum_{\alpha_k=1}^{N_k} J_k(\omega_k^{\alpha_k},t)[\rho],
\end{eqnarray}
where the superoperator $L_0$ and jump superoperator $J_k$ are
\begin{eqnarray}
\label{continuouspart}
 L_{0}(t)[\rho]&=&-\frac{i}{\hbar}[H(t),\rho]-\frac{1}{2}\sum_{k=1}^2  \sum_{\alpha_k=1 }^{N_k}r_k(\omega_k^{\alpha_k})\left\{A_k^\dag(\omega_k^{\alpha_k},t)A_k(\omega_k^{\alpha_k},t),\rho\right\},\\
\label{jumppart}
J_k(\omega_k^{\alpha_k},t)[\rho]&=&r_k(\omega_k)A_k(\omega_k^{\alpha_k},t)\rho A_k^\dag(\omega_k^{\alpha_k},t),
\end{eqnarray}
respectively. Applying the Dyson series to Eq.~(\ref{peturbedmastereq}), we obtain a formal solution:
\begin{eqnarray}
\label{Dysonformalsolutionrho}
\rho(t)&=&G_0(t,0)\left[\rho_0\right]+\nonumber \\
&& \sum_{M=1}^\infty\sum_{\overrightarrow{\omega}_M}    \left(\prod_{i=M}^1 \int_{0}^{t_{i+1}}dt_i \right)   G_0(t,t_M)J_{k_M}(\omega_{k_ M}^{\alpha_{k_M}},t_M)G_0(t_M,t_{M-1})\cdots \nonumber  \\
&&J_{k_1}(\omega_{k_1}^{\alpha_{k_1}},t_1)G_0(t_1,0)\left[\rho_0\right]\nonumber\\
&=&\sumint_M {\cal D}(t)\hspace{0.1cm} %\left(\prod dt_i\right)
G_0(t,t_M)J_{k_M}(\omega_{k_ M}^{\alpha_{k_M}},t_M) G_0(t_M,t_{M-1})\cdots J_{k_1}(\omega_{k_1}^{\alpha_{k_1}},t_1) G_0(t_1,0)\left[\rho_0\right],
\end{eqnarray}
where
\begin{eqnarray}
G_0(t,t')=T_{-}e^{\int_{t'}^{t}d\tau{L}_0(\tau)},
\end{eqnarray}
and we set $t_0=0$ and $t_{M+1}=t$. In the second summation of the above equation, the index
\begin{eqnarray}
\label{quantumjumptraj}
\overrightarrow{\omega}_M=\{\omega_{k_M}^{\alpha_{k_M}},\cdots,\omega_{k_1}^{\alpha_{k_1}}\}
\end{eqnarray}
is a sequence of $M$ chronologically ordered Bohr frequencies, where $M\ge 1$ is an arbitrary positive integer, $k_i$ ($i=M,\cdots,1$) is equal to $1$ or $2$ for the $k_i$-th heat bath, and the superscript $\alpha_{k_i}$ is equal to one of $1,\cdots,N_{k_i}$. This summation means that we sum over all possible $\overrightarrow{\omega}_M$, the number of which is $(N_1+N_2)^M$. In the second equation of Eq.~(\ref{Dysonformalsolutionrho}), shorthand notation is used to denote these integrals and summations with respect to all possible arrangements. The reader is reminded that all superoperators therein act on all operators on their right-hand sides.

Let the trace of the integrant of Eq.~(\ref{Dysonformalsolutionrho}) be $p(\overrightarrow{\omega}_M,t)$, i.e.,
\begin{eqnarray}
\label{probdensityfunctional}
p(\overrightarrow{\omega}_M,t)={\rm Tr}\left[G_0(t,t_M)J_{k_M}(\omega_{k_ M}^{\alpha_{k_M}},t_M) G_0(t_M,t_{M-1})\cdots J_{k_1}(\omega_{k_1}^{\alpha_{k_1}},t_1) G_0(t_1,0)\left[\rho_0\right]\right].
\end{eqnarray}
Dividing and multiplying the integrant by the functional, Eq.~(\ref{Dysonformalsolutionrho}) becomes
\begin{eqnarray}
\label{probabilityform}
\rho(t)%&=&\int_C {\cal D}(t)\hspace{0.1cm} \frac{ p\{\omega \}}{p\{\omega \}}
%G_0(t,t_M)J_{k_M}(\omega_{k_ M}^{\alpha_{k_M}},t_M) G_0(t_M,t_{M-1})\cdots J_{k_1}(\omega_{k_1}^{\alpha_{k_1}},t_1) %G_0(t_1,0)\left[\rho_0\right]\nonumber\\
&=&\sumint_M {\cal D}(t)\hspace{0.1cm}  p(\overrightarrow{\omega}_M,t ) {\widetilde \rho}(\overrightarrow{\omega}_M,t)  ,
\end{eqnarray}
where the normalized operator
\begin{eqnarray}
\label{densitymatrixquantumjumptraj}
%{\widetilde \rho}\{\omega_{k_j}\}=
{\widetilde \rho}(\overrightarrow{\omega}_M,t)=\frac{1 }{p(\overrightarrow{\omega}_M ,t)}
G_0(t,t_M)J_{k_M}(\omega_{k_ M}^{\alpha_{k_M}},t_M) G_0(t_M,t_{M-1})\cdots J_{k_1}(\omega_{k_1}^{\alpha_{k_1}},t_1) G_0(t_1,0)\left[\rho_0\right].
\end{eqnarray}
The following crucial step is to interpret $p(\overrightarrow{\omega}_M,t)$ as a probability density of observing the quantum trajectory $ \overrightarrow{\omega}_M $. There are two reasons for this interpretation. First, the functional is positive since $G_0$ and $J_k$ in Eqs.~(\ref{continuouspart}) and~(\ref{jumppart}) are Kraus superoperators and have a positivity property~\cite{Breuer2000,Liu2016a}. Second, this term is normalized, which we can see by taking traces on both sides of Eq.~(\ref{probabilityform}). Correspondingly, we call ${\widetilde \rho}(\overrightarrow{\omega}_M,t)$ the density matrix of the quantum trajectory $\overrightarrow{\omega}_M$. Therefore, Eq.~(\ref{probabilityform}) implies that the reduced density matrix $\rho(t)$ is equal to an average of ${\widetilde \rho}(\overrightarrow{\omega}_M,t)$ with respect to the probability density $p(\overrightarrow{\omega}_M,t)$.

The form of Eq.~(\ref{densitymatrixquantumjumptraj}) tells us how to achieve the history-dependent ${\widetilde \rho}(\overrightarrow{\omega}_M,t)$: given an initial density matrix $\rho_0$, under the action of the superpropagator $G_0(t,0)$, the matrix %Editor: Please ensure that the intended meaning has been maintained in this edit. Alternatively, please consider 'this superoperator'.
continuously evolves until time $t_1$ and is then acted on by the jump superoperator $J_{k_1}(\omega_{k_1}^{\alpha_{k_1}},t_1)$. Similar processes are repeated $M$ times. Finally, the continuous evolution of $G_0(t,t_M)$ ends the quantum trajectory. This scheme can be rigorously treated in a quantitative way. Because this context has been described comprehensively in textbooks~\cite{Breuer2002,Wiseman2010} and previous articles~\cite{Liu2016a,Liu2018}, we do not further explain the details.

\subsection{Stochastic heat and steady-state work }
The unravelling of the Floquet quantum master equation~(\ref{FQME}) into quantum trajectories is not just a formal manipulation. This unravelling process can be physically interpreted as the dynamic evolution of the quantum system with continuous measurements of the energy changes of the quantum heat baths~\cite{Breuer1997,Kist1999,Breuer2003,Liu2018}: $G_0(t_{i},t_{i-1})$ ($i=1,\cdots,M$) indicates that the measured energies of the two heat baths remain constant during the time interval $(t_i,t_{i-1})$, while $J_{k_i}(\omega_{k_i}^{\alpha_{k_i}},t_i)$ indicates that the energy of the $k_i$-th heat bath changes by an amount of $\hbar\omega_{k_i}^{\alpha_{k_i}}$ at time $t_i$. Importantly, from a thermodynamic point of view, these quanta can be regarded as the heat exchanged between the quantum system and the heat baths~\cite{Breuer2003,DeRoeck2006,Derezinski2008,Crooks2008,Horowitz2012,Hekking2013,Liu2014,Liu2014a,Liu2018,Manzano2015,Manzano2018}: if these quanta are positive, heat is released to the heat baths; otherwise, heat is absorbed from the heat baths. Based on this observation, given a quantum jump trajectory $\overrightarrow{\omega}_M$, we define the {\it net} heat released to the $k$-th heat bath $Q_k(\overrightarrow{\omega}_M)$ as being equal to a sum of the quanta $\hbar\omega_{k_i}^{\alpha_i}$ with special $k_i=k$, i.e.,
\begin{eqnarray}
\label{netheat}
Q_{k}(\overrightarrow{\omega}_M )&=&\sum_{i=1}^M\delta_{k,k_i }\hbar\omega_{k_i}^{\alpha_i} \nonumber \\
&=&Q_k^+(\overrightarrow{\omega}_M )+Q_k^-( \overrightarrow{\omega}_M ),
\end{eqnarray}
where $\delta$ is the Kronecker delta. In the second line of Eq~(\ref{netheat}), we additionally define $Q_k^{\pm}(\overrightarrow{\omega}_M )$, which represents the heat released to the $k$-th bath equal to the sums of the positive and negative quanta along the same quantum trajectory.

Another important thermodynamic quantity is work. Recently, there have been many controversies about the definition of work in a quantum regime (see the comprehensive review by B\"{a}umer et al.~\cite{Bumer2018} and the literature therein). In this paper, because we are concerned with the statistics of a quantum heat engine under long time limits, we propose a definition for {\it steady-state} work:
\begin{eqnarray}
\label{asywork}
W(\overrightarrow{\omega}_M )=-\left[Q_{1}(\overrightarrow{\omega}_M )+Q_{2}(\overrightarrow{\omega}_M )\right].
\end{eqnarray}
Because the dimension of the quantum system is assumed to be finite and the net heat is temporally extensive, %Editor: Please ensure that the intended meaning has been maintained in this edit.
this definition is consistent with the first law of thermodynamics. In addition, the Markov characteristics of the Floquet quantum master equation imply that the effects of the initial conditions of the quantum system disappear under long time limits. Hence, steady-state work shall have nothing to do with the controversy surrounding the measurement-based definition of quantum work.

Quantum trajectories are random events. Both heat and steady-state work are stochastic quantities. Hence, obtaining their distributions is crucial for studying stochastic heat engines. According to Eqs.~(\ref{netheat}) and~(\ref{asywork}), $Q_k^{\pm}$ represents the most obvious fundamental quantities. Let the joint probability distribution of a heat vector $\overrightarrow{Q}=(Q_1^+,Q_1^-,Q_2^+,Q_2^-)$ be
\begin{eqnarray}
\label{probQvector}
p(\overrightarrow{ Q}) =\sumint_M {\cal D}(t)\hspace{0.1cm}  p(\overrightarrow{\omega}_M,t ) \prod_{k=1}^2\delta(Q_k^+-Q_k^{+}(\overrightarrow{\omega}_M) ) \delta(Q_k^--Q_k^{-}(\overrightarrow{\omega}_M)).
\end{eqnarray}
There are two methods to obtain this distribution~\cite{Liu2016a,Liu2018}. One is to simulate the quantum trajectories and to construct the histogram of the heat. An alternative method is to compute the characteristic function of the distribution, i.e.,
\begin{eqnarray}
\label{heatCF}
\Phi\left(\overrightarrow{ \chi}\right)&=&\int \left(\prod_{k=1}^2 dQ_k^+ dQ_k^- \right) p\left(\overrightarrow {Q}\right) e^{i \overrightarrow {\chi} \cdot\overrightarrow {Q}  } ={\rm Tr}[\hat\rho(t)],
%\\
%\int e^{i\eta w}p(w)dw&=&\int e^{i\eta w}dw\int \delta (w+q_1+q_2)p(q_1,q_2)dq_1dq_2\\
%&=&\int e^{-i\eta(q_1+q_2)}p(q_1,q_2)dq_1dq_2\nonumber\\
%&=&\Phi(-\eta,-\eta) \nonumber
\end{eqnarray}
where the parameters $\overrightarrow{\chi}=(\chi_1^+,\chi_1^-,\chi_2^+,\chi_2^-)$; then, the inverse Fourier transform of the function is performed. Therein, we introduce a characteristic operator $\hat\rho(t)$ after substituting Eqs.~(\ref{probdensityfunctional}) and~(\ref{probQvector}) into Eq.~(\ref{heatCF}). It is not difficult to prove that this operator satisfies an equation analogous to the Floquet quantum master equation~(\ref{FQME}):
\begin{eqnarray}
\label{equationofmotionforrhohat}
\partial_t \hat{\rho}(t)&=&-\frac{i}{\hbar}[H(t), \hat{{\rho}}]+D_1(t,\chi_1^+,\chi_1^-)[\hat{\rho}(t)]+D_2(t,\chi_2^+,\chi_2^-)[\hat{\rho}(t)],
\end{eqnarray}
where
\begin{eqnarray}
\label{modifieddissipation}
D_k(t,\chi_k^+,\chi_k^-)[\hat\rho ]&=&\sum_{\omega_k^{\alpha_k}>0 }  r_k (\omega_k^{\alpha_k})\left[ e^{i\chi_k^+ \hbar\omega_k^{\alpha_k} }A_k(\omega_k^{\alpha_k},t)\hat\rho  A_k^\dag(\omega_k^{\alpha_k},t)- \frac{1}{2}
\left\{A_k^\dag (\omega_k^{\alpha_k},t)A_k(\omega_k^{\alpha_k},t),\hat\rho \right\}\right]\nonumber \\
&&+\sum_{\omega_k^{\alpha_k}<0 } r_k (\omega_k^{\alpha_k})\left[ e^{i\chi_k^-\hbar\omega_k^{\alpha_k}} A_k(\omega_k^{\alpha_k},t)\hat\rho  A_k^\dag(\omega_k^{\alpha_k},t)- \frac{1}{2}
\left\{A_k^\dag (\omega_k^{\alpha_k},t)A_k(\omega_k^{\alpha_k},t),\hat\rho \right\}\right].\nonumber \\
%&&\sum_{\omega>0}r_k(-\omega)\left[A_k^\dag(\omega,t)O A_k(\omega,t)- \frac{1}{2}
%\left\{A_k(\omega,t)A_k^\dag(\omega,t),O\right\}\right].
\end{eqnarray}
Indeed, if $\overrightarrow{\chi}=0$, these equations are identical. Appendix A provides some details of the above.
We also note that if $\chi_k^+=\chi_k^-$, Eq.~(\ref{equationofmotionforrhohat}) reduces to the modified quantum master equation presented by Cuetara et al.~\cite{Cuetara2015}. Their paper was concerned with the statistics of the net heat $Q_k$ instead of the statistics of $Q_k^{\pm}$. In contrast, if we follow their idea to derive Eq.~(\ref{modifieddissipation}), we would again carry out two energy measurement schemes on the heat baths and repeat a lengthy derivation analogous to the derivation of the Floquet quantum master equation~(\ref{FQME})~\cite{Gasparinetti2014,Esposito2009}. Therefore, the quantum trajectory shows its flexibility when we are exploring the new statistics. It is worth emphasizing that these two methods of computing the heat distributions are complementary. The simulation method is straightforward and is also useful when computing quantum systems with larger degrees of freedom~\cite{Breuer2002}. The characteristic function method needs to solve differential equations and is superior in analysis, e.g., when investigating the large deviation principle, which will be shown shortly. Finally, we want to point out Eq.~(\ref{equationofmotionforrhohat}) gives the ensemble-averaged formulas of the currents of the net heat~\cite{Liu2016a,Liu2018}:
\begin{eqnarray}
\label{currentformual}
%&=&\partial_{\chi_k} \Phi(\chi_1,\chi_2)|_{\chi_1=\chi_2=0} \nonumber \\
J_k(t)&=&\sum_{\alpha_k=1 }^{N_k} \hbar\omega_k^{\alpha_k} r_k(\omega_k^{\alpha_k}){\rm Tr}\left[ A_k(\omega_k^{\alpha_k},t)\rho(t) A_k^\dag(\omega_k^{\alpha_k},t)
\right ] \nonumber \\
&=&J_k^+(t) + J_k^-(t).
%\sum_{\omega_{t_1}} \hbar\omega  \gamma_k(\omega){\rm Tr}[  A(\omega_{t_1},t_1)A^\dag (\omega_{t_1},t_1)]
\end{eqnarray}
In the second line of Eq.~(\ref{currentformual}), $J_k^+(t)$ ($J_k^-(t)$) is the ensemble average of the heat current released to (absorbed from) the $k$-th heat bath, and the expressions of these ensemble averages are the same as the expressions in the first line except that their sums are with respect to the terms with positive (negative) Bohr frequencies.

\subsection{Two stochastic efficiencies}
As an application of the stochastic theory for the Floquet quantum heat engine, we focus on two types of stochastic efficiencies. One is~\cite{Verley2014a,Verley2014b,Manikandan2019,Gingrich2014,Jiang2015,Proesmans2015}
\begin{eqnarray}
\label{effdef1}
\eta_s(\overrightarrow{\omega}_M)=-\frac{1}{\eta_C }\frac{W(\overrightarrow{\omega}_M)}{Q_1(\overrightarrow{\omega}_M)},
\end{eqnarray}
where $\eta_C=1-\beta_1/\beta_2$ is the Carnot efficiency. The subscript $s$ denotes that the net heat is absorbed from the single hot heat bath. The other efficiency is
\begin{eqnarray}
\label{effdef2}
\eta_d (\overrightarrow{\omega}_M)=-\frac{1}{\eta_C }\frac{W(\overrightarrow{\omega}_M)}{Q^-(\overrightarrow{\omega}_M)},
\end{eqnarray}
where
\begin{eqnarray}
Q^-( \overrightarrow{\omega}_M )=Q_1^-(\overrightarrow{\omega}_M)+Q_2^-(\overrightarrow{\omega}_M)
\end{eqnarray}
is the absorbed (negative) heat from the two heat baths along a quantum trajectory $\overrightarrow{\omega}_M $. To distinguish the former efficiency, we add the subscript $d$ to denote that the heat originates from both heat baths. Obviously, along the same trajectory, $\eta_s$ is always greater than or equal to $\eta_d$.

Compared with $\eta_s$, which has attracted considerable interest~(see \cite{Manikandan2019} and references therein), to the best of our knowledge, fewer studies have been conducted on $\eta_d$. Strictly speaking, the latter is closer than the former to the principle of defining the efficiency of a heat engine since the steady-state work~(\ref{asywork}) includes the heat contributions from both the hot heat bath and the cold heat bath. This subtle distinction is absent in macroscopic heat engines. According to the second law of thermodynamics~\cite{Szczygielski2013},
\begin{eqnarray}
\label{secondlaw}
\beta_1 J_1 + \beta_2 J_2\ge 0,
\end{eqnarray}
we can prove that the Carnot efficiency is the upper bound of the ensemble averages of these two efficiencies, which are
\begin{eqnarray}
\label{maxefficiency2}
\overline{\eta}_s&=&-\frac{\langle W\rangle }{\langle Q_1\rangle }=\frac{1}{\eta_C }\left( 1+\frac{J_2  }{  J_1 }\right),
\end{eqnarray}
and
\begin{eqnarray}
\overline{\eta}_d&=&-\frac{\langle W\rangle }{\langle Q^-\rangle }=\frac{1}{\eta_C}\left( 1+\frac{J^+  }{  J^- }\right),
\end{eqnarray}
where $J^\pm=J^\pm_1+J_2^\pm$: $\overline{\eta}_s$ is always greater or equal to $\overline{\eta}_d$, while $\overline{\eta}_s \le  \eta_C$ is a straightforward consequence of Eq.~(\ref{secondlaw}). The second law of thermodynamics also implies an additional inequality:
\begin{eqnarray}
\label{maxefficiency2}
 \overline{\eta}_d &\le& 1-\frac{\beta_-}{\beta_+},
\end{eqnarray}
where $\beta_{\pm}$ represents the current-dependent ``temperatures'', which are defined as
\begin{eqnarray}
\beta_\pm=\frac{1}{J^\pm}\left(J_1^\pm {\beta_1} +J_2^\pm {\beta_2}\right).
\end{eqnarray}
Because we are interested in the steady-state situation, we do not explicitly write the time parameters in these currents. Further considering that the quantum thermal machine operates in heat engine mode, i.e., $\langle W\rangle > 0$, we always have
\begin{eqnarray}
\beta_1\le \beta_- < \beta_+\le \beta_2.
\end{eqnarray}
Obviously, this result leads to $\overline{\eta}_d \le  \eta_C$ again.

Analogous to the case with heat, the distributions of the efficiencies can be computed by two methods. Because Eqs.~(\ref{effdef1}) and~(\ref{effdef2}) have the same structure, we express these distributions in a unified form:
\begin{eqnarray}
\label{probeta1}
p(\eta)&=&\sumint_M {\cal D}(t)\hspace{0.1cm} p(\overrightarrow{\omega}_M,t ) \delta(\eta-\eta(\overrightarrow{\omega}_M))\\
\label{probeta2}
&=& \int dq_a dq_b p(q_a,q_b) \delta\left(\eta+\frac{q_b}{q_a}  \right),
\end{eqnarray}
where $q_a=\eta_CQ_1$ or $\eta_C Q^-$ depends on which efficiency is studied and $q_b=W$. The first equation indicates that the distributions can be constructed by simulating a sufficient number of quantum trajectories. The second equation needs to first solve the distribution $p(q_a,q_b)$, which can be obtained from the distribution $p(\overrightarrow{Q})$ (see Eq.~(\ref{probQvector})). However, a more intriguing component is the large deviation function of $p(\eta)$~\cite{Touchette2008}:
\begin{eqnarray}
\label{LDFdefinition}
I(\eta)=-\lim_{t\rightarrow \infty} \frac{1}{t}\ln p(\eta).
\end{eqnarray}
Obviously, this function may also be computed by a direct simulation of the quantum trajectories; however, this approach will be impractical if one wants the full profile of $I(\eta)$. Importantly, Verley et al.~\cite{Verley2014b} presented an alternative method that is based on the scaled cumulant generation function $\widetilde{\phi}(\chi_a,\chi_b)$ of the distribution $p(q_a,q_b)$:
\begin{eqnarray}
\label{computingLDF}
I(\eta)&=&-\min_{\chi_b} \widetilde{\phi}(\chi_b\eta, \chi_b).
\end{eqnarray}
It is easy to see that for the two efficiencies, the concrete expressions for the scaled cumulant generation functions are
\begin{eqnarray}
\label{computingLDF1}
\widetilde{\phi}_s(\chi_b\eta, \chi_b)&=&  \phi(\chi_b(\eta\eta_C-1),\chi_b(\eta\eta_C-1),-\chi_b,-\chi_b), \\
\label{computingLDF2}
\widetilde{\phi}_d(\chi_b\eta, \chi_b)&=& \phi(-\chi_b,\chi_b(\eta\eta_C-1),-\chi_b,\chi_b(\eta\eta_C-1)),
\end{eqnarray}
where $\phi(\overrightarrow{\chi})$ on the right-hand sides is the scaled cumulant generation function for the heat vector $\overrightarrow{Q}$:
\begin{eqnarray}
\label{scaledcgf}
\phi(\overrightarrow{\chi})=\lim_{t\rightarrow \infty} \frac{1}{t}\ln \Phi(-i\overrightarrow{\chi}).
\end{eqnarray}
Again, we add the subscripts $d$ and $s$ to $\widetilde{\phi}$ to distinguish between these two types of stochastic efficiencies.

\section{Two-level Floquet quantum heat engine}
\label{section4}
In this section, we use a two-level quantum system~\cite{Breuer1997,Kohler1997,Szczygielski2013,Langemeyer2014,Gasparinetti2014,Cuetara2015} as the quantum working medium to illustrate the general results presented above. The periodically driven Hamiltonian is
\begin{eqnarray}
\label{TLS}
H(t)=\frac{1}{2} {\hbar\omega_0}\sigma_z +\frac{1}{2}{\hbar\Omega_R}\left(\sigma_+ e^{-i\Omega t}+\sigma_- e^{i\Omega t}\right),
\end{eqnarray}
where $\omega_0$ is the transition frequency of the bare system, $\Omega_R$ is the Rabi frequency, and $\Omega$ is the frequency of the periodic external field. The Floquet bases and the quasi-energies of the quantum system are
\begin{eqnarray}
\label{TLSFloquetbases}
|u_{\pm}(t)\rangle =\frac{1}{\sqrt{2\Omega'}}
\left(\begin{array}{c}
 \pm \sqrt{\Omega'\pm\delta}\\
 e^{i\Omega t}\sqrt{\Omega'\mp\delta},
\end{array}\right),
\end{eqnarray}
and
\begin{eqnarray}
\label{TLSquasienergies}
\epsilon_\pm=\frac{\hbar}{2}(\Omega \pm \Omega'),
\end{eqnarray}
respectively, where $\Omega'=\sqrt{\delta ^2+\Omega_R^2}$ and the detuning parameter $\delta=\omega_0-\Omega$.
We assume that the couplings between the quantum system and the hot and cold heat baths are transverse ($\sigma_x$-coupling) and longitudinal ($\sigma_z$-coupling)~\cite{Szczygielski2013}, respectively. There are six Lindblad operators for the system and the hot heat bath: three of them (with Bohr frequencies $\Omega $, $(\Omega-\Omega')$, and $(\Omega+\Omega')$) are
\begin{eqnarray}
\label{Lindbladoperators1}
A_1(\Omega,t)&=&\frac{\Omega}{2\Omega'}\left(|u_+(t)\rangle\langle u_+(t)|-|u_-(t)\rangle\langle u_-(t)| \right)e^{-i\Omega t},\nonumber\\
A_1(\Omega -\Omega',t)&=&\left(\frac{\delta-\Omega'}{2\Omega'} \right)|u_+(t)\rangle\langle u_-(t)|e^{-i\Omega t},\\
A_1(\Omega+\Omega',t)&=&\left(\frac{\delta+\Omega'}{2\Omega'} \right) |u_-(t)\rangle\langle u_+(t)| e^{-i\Omega t}.\nonumber
\end{eqnarray}
The other three Lindblad operators $A_k(\omega,t)$ (with $\omega=-\Omega $, $-(\Omega -\Omega')$, and $-(\Omega +\Omega')$) are the adjoint operators of Eq.~(\ref{Lindbladoperators1}). There are a total of three Lindblad operators for the system and the cold bath:
\begin{eqnarray}
\label{Lindbladoperators2}
A_2(0,t)&=&\frac{\delta}{\Omega'}\left(|u_+(t)\rangle\langle u_+(t)|-|u_-(t)\rangle\langle u_-(t)| \right),\nonumber\\
A_2(\Omega',t)&=&-\frac{\Omega_R }{\Omega'}|u_-(t)\rangle\langle u_+(t)|, \\
A_2(-\Omega',t)&=&-\frac{\Omega_R }{\Omega'}|u_+(t)\rangle\langle u_-(t)|.
\end{eqnarray}
The reason that we do not use the same type of coupling between the quantum system and both heat baths is that such a thermal machine cannot operate as a heat engine. A detailed explanation is provided in Appendix B. We assume that the Fourier transformations of the correlation functions $r_k(\omega)={\cal A}|{\omega}|^3 {\cal N}_k(\omega)$ for $\omega<0$; otherwise, $r_k(\omega)={\cal A}|{\omega}|^3 [{\cal N}_k(\omega)+1]$~\cite{Breuer2000}, where
\begin{eqnarray}
{\cal N}_k(\omega)=\frac{1}{e^{\beta_k \hbar|\omega| }-1}.
\end{eqnarray}
The coefficient $A$ represents the strength of the coupling between the quantum system and the heat baths.

\begin{figure}
\includegraphics[width=1.\columnwidth]{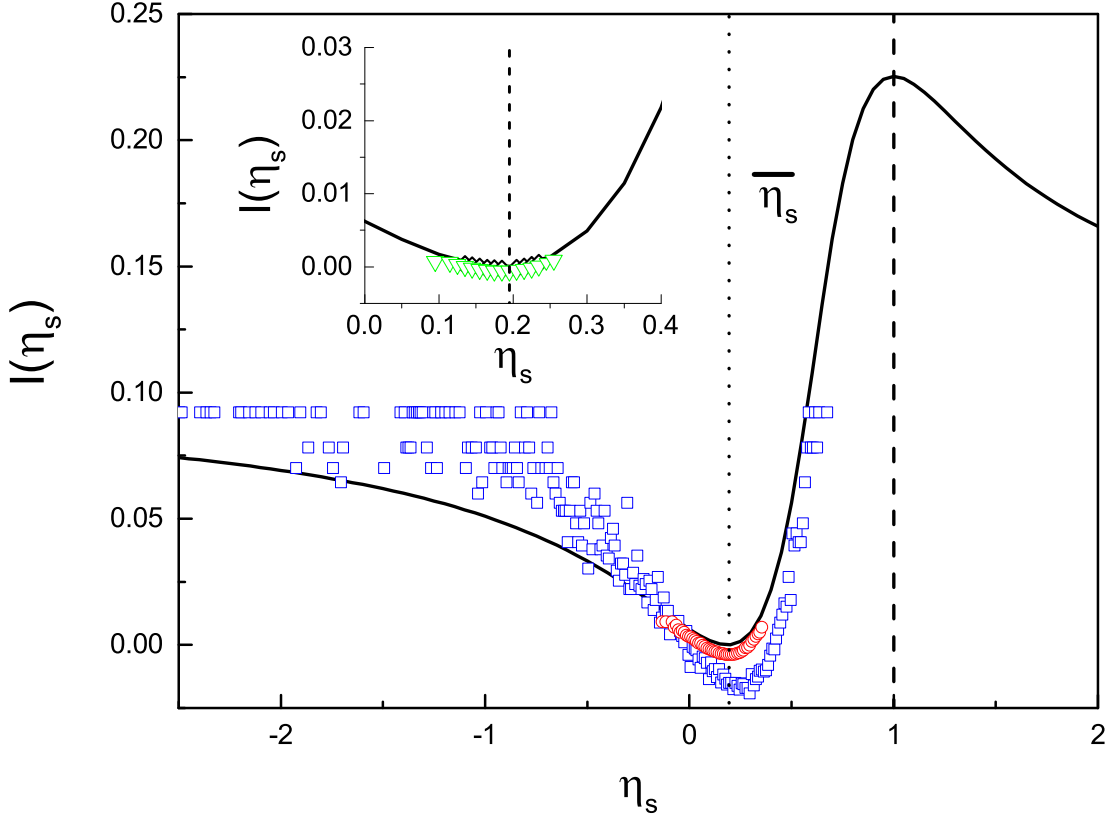}
\caption{The large deviation functions $I(\eta_s)$ of the stochastic efficiency $\eta_s$ for the quantum two-level engine. The parameters used are $\omega_0=1$, $\Omega_R=1.8$, $\Omega=0.9$ (red tune),
${\cal A} = 1$, $\beta_1 = 1/3$, and $\beta_2 = 1$. The durations of simulations are $50$ (squares), $500$ (circles), $5\times 10^3$ (inverted triangles in the inset), and $10^4$ (upright triangles in the inset). The number of quantum trajectories for each simulation is $10^4$. The solid curve is the large deviation function %Editor: Abbreviations and acronyms are typically defined the first time the term is used within the main text and then used throughout the remainder of the manuscript. Please consider adhering to this convention.
obtained by numerically solving the characteristic function and using Eq.~(\ref{computingLDF}). }
\label{fig2}
\end{figure}

Our interest is to compute the large deviation functions of the stochastic efficiencies. Figure~(\ref{fig2}) shows the results of the efficiency $\eta_s$ obtained by simulating the quantum trajectories under a set of parameters. Figure 1(b) shows a simulated quantum trajectory and the absorbed and released heat along it. All computational details are given in Appendix D of~\cite{Liu2016a}, and thus, we do not explain the details here. We note that as the simulation time increases, these data quickly converge to a fixed curve and shrink to a small range, where they have almost the same minima. The narrowing ranges of the nonzero large deviation functions are not surprising since larger simulation times result in smaller fluctuations of the net heat current flowing out (into) the hot (cold) heat bath. According to the universal theory of stochastic efficiency~\cite{Verley2014b}, the minimum of the large deviation function is at the ensemble-averaged efficiency $\overline{\eta}_s$ or at the most probable efficiency. We then compute the average and depict it in Fig.~(\ref{fig2}) (see the dotted lines therein). We see that the theoretical prediction and our simulation agree very well. In addition, the above mentioned universal theory also predicts that there is a maximum of the large deviation function at the Carnot efficiency or at the least probable efficiency $\eta_s=1$. However, the simulated data do not show this maximum.

To obtain the full profile of the large deviation function of the stochastic efficiency, we solve the characteristic function $\Phi(\overrightarrow{\chi})$ and then obtain $I_s(\eta)$ through Eq.~(\ref{scaledcgf}). The computational procedure is standard. First, we write the characteristic operator $\hat\rho$ in a Pauli matrix-like representation constructed by the Floquet basis, i.e.,
\begin{eqnarray}
\label{chap6hatrhoA}
\hat{\rho}(t)&=&\frac{ n_e(t)+ n_g(t)}{2}I+\frac{n_e(t)-n_g(t)}{2}\sigma'_z(t)+u(t)\sigma'_+(t)+ v(t)\sigma'_-(t),
\end{eqnarray}
where $I$ is the identity operator (for the definitions of $\sigma_z'(t)$ and $\sigma_\pm'(t)$, see Appendix B). Then, we have
\begin{eqnarray}
\Phi(\overrightarrow{\chi})=n_e(t)+n_g(t).
\end{eqnarray}
The ``population'' vector $\overrightarrow{ n}=(n_e,n_g)^T$ satisfies a set of ordinary differential equations:
\begin{eqnarray}
\frac{d}{dt}\overrightarrow{n}=\textbf{A}(\overrightarrow\chi)\overrightarrow{n},
\end{eqnarray}
where the matrix elements $(\textbf{A})_{i,j=1}^2$ are $\overrightarrow{\chi}$-dependent constants. These equations can be easily solved, and we obtain the scaled cumulant generation function
\begin{eqnarray}
\phi(\overrightarrow {\chi} )= \lambda_+(-i\overrightarrow\chi),
\end{eqnarray}
where $\lambda_+$ represents the larger eigenvalue of the matrix $\textbf{A}(\overrightarrow{\chi})$. Its concrete expression is provided in Appendix C. Figure~(\ref{fig2}) shows the data computed by Eq.~(\ref{computingLDF}) (see the solid curve therein). We see that the exact numerical results agree with the simulated large deviation functions around $\overline{\eta}_s$. In particular, a maximum at $\eta_s=1$ appears (see the dashed line). Although the engine is quantum, in the Floquet basis, its dynamic component relevant to stochastic thermodynamics is identical to a classic rate process. Hence, the good agreement between our quantum trajectory simulation and the universal theory of stochastic efficiency developed by Verley et al.~\cite{Verley2014a} is not very surprising.

We can carry out an analogous analysis for the uncommon efficiency $\eta_d$. We find that the results of the quantum trajectory simulation are similar to those of the conventional efficiency $\eta_s$. Hence, we resort to the characteristic function method again. As predicted by the powerful universal theory, the most probable efficiency is precisely located at the ensemble-averaged efficiency $\overline{\eta}_d$. Note that this average is six times smaller than the average of $\eta_s$ under the same set of parameters. Intriguingly, the large deviation function of this efficiency has no maximum at the Carnot efficiency. To explain this observation, we reexamine the assumptions used in the theory of Verley et al.~\cite{Verley2014a,Verley2014b}. We note that although the joint probability distribution of $\wp=W/t$ and $j^- = Q^-/t$ still follows the large deviation principle, it does not satisfy any fluctuation theorems. In fact, contrary to the stochastic variable $j_1= Q_1/t$, which may be positive or negative, the stochastic variable $j^-$ is always negative. Therefore, no fluctuation theorems are involved in such a situation. In Appendix C, we present several comments on the fluctuation theorems for the stochastic Floquet quantum engine.
\begin{figure}
\includegraphics[width=1.\columnwidth]{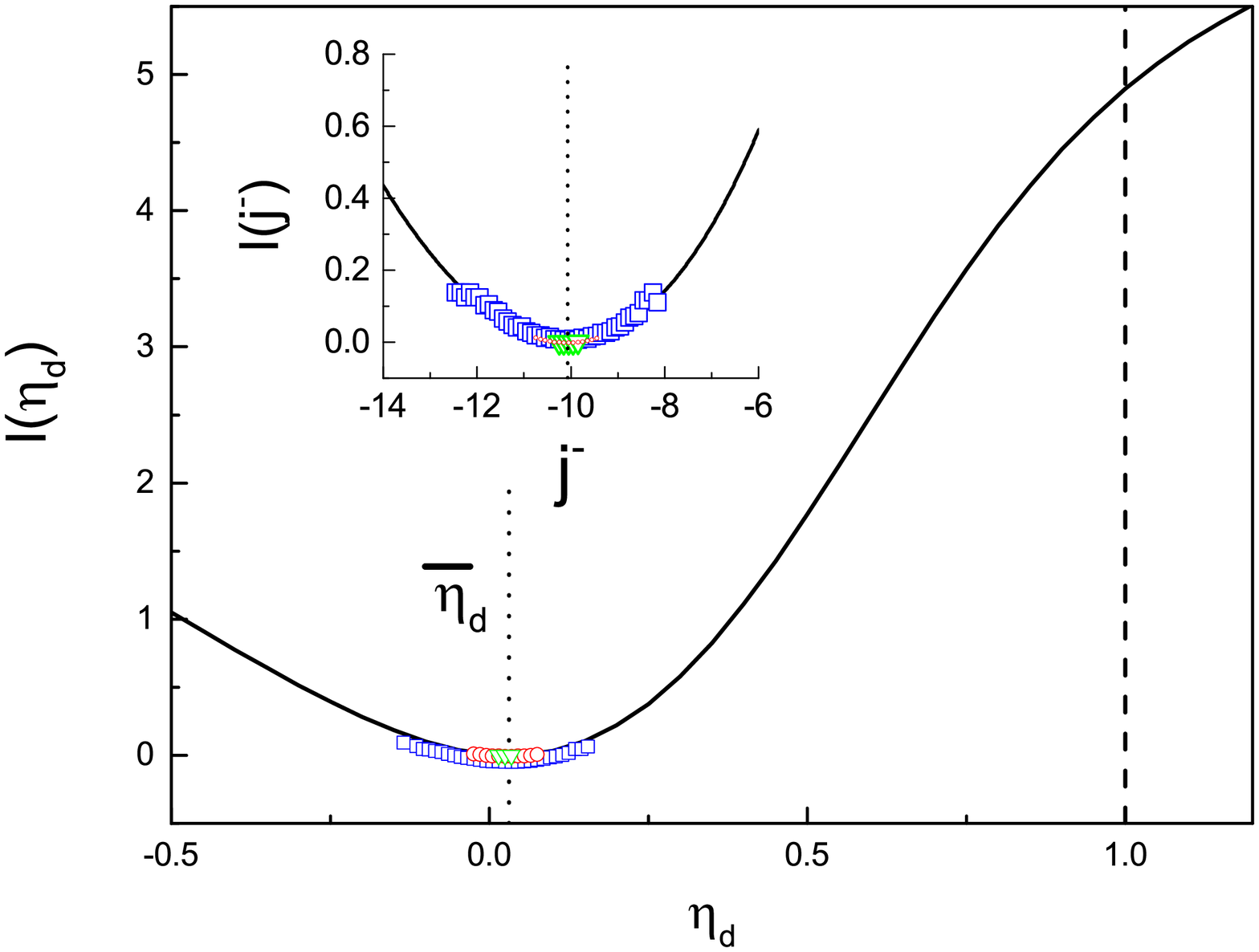}
\caption{The large deviation functions $I(\eta_d)$ of the stochastic efficiency $\eta_d$ for the quantum two-level engine. The parameters and the meanings of the solid curves, the dashed and dotted lines, and the various symbols are the same as those in Fig.~(\ref{fig2}). The inset is the large deviation function $I(j^-)$ of the absorbed total heat current $j^-$, which we compute by performing a Legendre transformation of the scaled cumulant generation function ${\phi}(0,\chi,0,\chi)$. The dotted line therein is the ensemble average $J^-$. }
\label{fig3}
\end{figure}

\section{Conclusion}
\label{section5}
In this paper, using the notion of quantum trajectory, we develop a stochastic framework for quantum heat engines described by the Floquet quantum master equation. We apply this framework to investigate the statistics of the two types of stochastic efficiencies that are defined along individual quantum trajectories. These different efficiencies originate from different supplies of heat: one is the net heat absorbed from the hot heat bath, and the other is the total heat absorbed from both heat baths. We find that although the ensemble averages of these two efficiencies follow the same upper bound of the Carnot efficiency, their fluctuation characteristics at the Carnot efficiency are distinct. The reason for this distinction is attributed to the lack of fluctuation theorems surrounding the joint probability distribution of the output power and the total negative heat current. There are several questions that need further study. One is to determine which efficiency can more optimally reflect the working performance of stochastic quantum heat engines. The second is the feasibility of computing the large deviation functions of the stochastic efficiencies by quantum trajectory simulations instead of the characteristic function method. This additional work will be essential if we address quantum systems with larger degrees of freedom. The last and perhaps more interesting question is how quantum phenomena affect the efficiency fluctuations, e.g., coherence and entanglement. Although quantum effects are highly concerned in studies of performance of quantum thermal machines~\cite{Levy2018}, their roles in the efficiency fluctuations were rarely mentioned. Several recent works have shown this possibility~\cite{Miller2019}. We hope to report these results in the near future.

\begin{acknowledgments}
This work was supported by the National Science Foundation of China under Grant No. 11174025 and No. 11575016. We also appreciate the support of the CAS Interdisciplinary Innovation Team (No. 2060299).
\end{acknowledgments}

\appendix
\iffalse
\section{Evolution equation of stochastic density matrix }
The density matrix of quantum jump trajectory ${\widetilde \rho}\{\vec\omega,t\}$ of Eq.~(\ref{densitymatrixquantumjumptraj}) can be achieved by the following procedures~\cite{Breuer2000}. Given an initial density matrix $\rho_0$. Evolving the density matrix with a continuous equation
\begin{eqnarray}
\label{continuouseq}
\partial_s \widetilde \pi(s)=L_0(s)[\widetilde \pi(s) ]
\end{eqnarray}
with the initial condition $\widetilde \pi(s=t_{i-1})=\widetilde \rho(t_{i-1})$.
\begin{eqnarray}
\widetilde\rho(s)=\widetilde \pi(s)/{\rm Tr}[\widetilde\pi(s)].
\end{eqnarray}
during the time interval $(t_{i},t_{i-1})$, the evolution equation is

At time $t_i$, a jump occurred
\begin{eqnarray}
\label{jumpeq}
\widetilde \rho(t_i)= \frac{1}{{\rm Tr}[A_k(\omega_k^{\alpha_k},t_{i})\widetilde\pi(t_{i}) A_k^\dag(\omega_k^{\alpha_k},t_{i}) ]}A_k(\omega_k^{\alpha_k},t_{i})\widetilde\pi(t_{i}) A_k^\dag(\omega_k^{\alpha_k},t_{i})
%J_k(\omega_k^{\alpha_k},t)[\widetilde \rho],
\end{eqnarray}
the probability of jump is
\begin{eqnarray}
r_k(\omega_k^{\alpha_k}) {\rm Tr}[A_k(\omega_k^{\alpha_k},t_{i})\widetilde\pi(t_{i}) A_k^\dag(\omega_k^{\alpha_k},t_{i}) ]
\end{eqnarray}
\fi

\section{Derivation of Eq.~(\ref{equationofmotionforrhohat})}
Substituting Eqs.~(\ref{probdensityfunctional}) and~(\ref{probQvector}) into Eq.~(\ref{heatCF}), we have
\begin{eqnarray}
\label{CFheatdef}
\Phi(\overrightarrow {\chi})%&=&\left\langle \exp(i\xi Q)\right\rangle\nonumber\\
&=&\sumint_M {\cal D}(t)\hspace{0.1cm}  p(\overrightarrow\omega,t)    e^{ i\overrightarrow {\chi} \cdot\overrightarrow {Q}(\overrightarrow{\omega})} \nonumber \\
&=&{\rm Tr}\left[\sumint_M {\cal D}(t)\hspace{0.1cm} G_0(t,t_M)e^{i\chi_{k_M}^{s_M}\hbar\omega_{k_ M}^{\alpha_{k_M}}}
J_{k_M}(\omega_{k_ M}^{\alpha_{k_M}},t_M) G_0(t_M,t_{M-1})\right.\nonumber\\
&&\left.\cdots e^{i\chi_{k_1}^{s_1}\hbar\omega_{k_1}^{\alpha_{k_1}}} J_{k_1}(\omega_{k_1}^{\alpha_{k_1}},t_1) G_0(t_1,0)\left[\rho_0\right]\right],
\end{eqnarray}
where we use the symbol $s_M$ ($M= N,\cdots,1$) to denote the sign of $\omega_{k_ M}^{\alpha_{k_M}}$: if the Bohr frequency is positive, $s_M=+$; otherwise, $s_M=-$. We immediately find that the entire term in the square brackets is almost the same as the formal solution of $\rho(t)$ (see Eq.~(\ref{Dysonformalsolutionrho})). The only difference is that each superoperator $J_k(\omega_k^{\alpha_k},t)$ is multiplied by a factor of either $\exp(i\chi_k^+\hbar\omega_k^{\alpha_k})$ or $\exp(i\chi_k^-\hbar\omega_k^{\alpha_k})$ depending on the sign of $\omega_k^{\alpha_k}$. Therefore, the characteristic operator $\hat{{\rho}}(t)$ satisfies Eq.~(\ref{equationofmotionforrhohat}).

\section{Conditions of a two-level quantum system as a Floquet quantum heat engine  }
Because the Floquet basis is complete and orthogonal, it shall be convenient to introduce Pauli matrix-like operators:
\begin{eqnarray}
\sigma_z'(t)&=&|u_+(t)\rangle\langle u_+(t)|-|u_-(t)\rangle\langle u_-(t)|, \nonumber \\
\sigma_{\pm}'(t)&=&|u_\pm(t)\rangle\langle u_\mp (t)|.
\end{eqnarray}
In addition, we define the following rates:
\begin{eqnarray}
\label{coeffs}
\Gamma^k_\Omega&=&\left(\frac{\Omega_R}{2\Omega'}\right)^2 r_k(\Omega), \nonumber \\%\left(\frac{\Omega_R}{2\Omega'}\right)^2 r_1(\Omega) \nonumber \\
\Gamma^k_{\Omega-\Omega'} &=&\left(\frac{\delta-\Omega_R}{2\Omega'}\right)^2 r_k(\Omega-\Omega'),\nonumber \\
\Gamma^k_{\Omega+\Omega'}&=&\left(\frac{\delta+\Omega_R}{2\Omega'}\right)^2 r_k(\Omega+\Omega'),\nonumber \\
\Gamma^k_{\Omega'}&=& \left(\frac{\Omega_R}{\Omega'}\right)^2 r_k(\Omega')
\end{eqnarray}
($k=1,2$). $\Gamma^k_{-\Omega} $, $\Gamma^k_{-(\Omega-\Omega')}$, $\Gamma^k_{-(\Omega+\Omega')}$, and $\Gamma^k_{-\Omega'}$ are analogously defined. We do not write them explicitly since they satisfy the KMS condition~(\ref{KMScondition}) with respect to the ones listed above.

We have mentioned that if a two-level quantum system interacts with two heat baths with the same interaction modes, the quantum machine cannot operate as a heat engine even if the two baths have different temperatures. To prove this statement, we need the concrete formulas for the heat current $J_k$ ($k=1,2$). For the case with identical $\sigma_x$-coupling, according to Eq.~(\ref{currentformual}), the currents are
\begin{eqnarray}
\label{currentTLSsigmax}
J_k=\hbar\Omega\left( \Gamma^k_\Omega-\Gamma^k_{-\Omega}  \right) &+&\hbar(\Omega+\Omega') \left( \Gamma^k_{\Omega+\Omega'}n_e-\Gamma^k_{-(\Omega+\Omega')}n_g \right) +  \hbar(\Omega-\Omega') \left( \Gamma^k_{\Omega-\Omega'}n_g-\Gamma^k_{-(\Omega-\Omega')}n_e\right),
\end{eqnarray}
where $n_e$ and $n_g$ are the populations of the quantum system occupying the Floquet bases  $|u_\pm(t)\rangle$. %Editor: Please ensure that the intended meaning has been maintained in this edit.
Eq.~(\ref{currentTLSsigmax}) also explains why the coefficients in Eq.~(\ref{coeffs}) are called rates. Because we are interested in long time limits, these populations are constants and can be obtained from the quantum master equation~(\ref{FQME}) in the steady state:
\begin{eqnarray}
\label{populationssigmax}
\left(\Gamma^1_{-(\Omega-\Omega')} + \Gamma^1_{\Omega+\Omega'} + \Gamma^2_{-(\Omega-\Omega')} + \Gamma^2_{\Omega+\Omega'} \right) n_e
=\left(\Gamma^1_{\Omega-\Omega'} + \Gamma^1_{-(\Omega+\Omega')} + \Gamma^2_{\Omega-\Omega'} + \Gamma^2_{-(\Omega+\Omega')} \right) n_g,
\end{eqnarray}
and $n_g+n_e=1$. As a quantum heat engine, we must require the output steady-state power to be positive, that is, $J_1+J_2<0$. Substituting the population solutions of Eq.~(\ref{populationssigmax}) and using the KMS condition~(\ref{KMScondition}), if $\Omega\ge\Omega'$, we easily find that the power is always negative. If $\Omega<\Omega'$, we can still prove that the power is negative; in contrast to the former case, the proof is slightly complex, and we have to use the original formulas of these rates. Considering that this is elementary, we do not show this proof here. For the other case with identical $\sigma_z$-coupling, the heat currents are
\begin{eqnarray}
\label{currentTLSsigmaz}
J_k=\hbar\Omega'\left(\Gamma^k_{\Omega'}n_e-\Gamma^k_{-\Omega'}n_g\right),
\end{eqnarray}
and the populations satisfy
\begin{eqnarray}
\left(\Gamma^1_{\Omega'} + \Gamma^2_{\Omega' }  \right) n_e
=\left(\Gamma^1_{-\Omega'} + \Gamma^2_{-\Omega'}  \right) n_g
\end{eqnarray}
and $n_g+n_e=1$. We immediately find that due to $J_1+J_2=0$, this quantum machine is unable to output steady-state power.

If the couplings are different with respect to the heat baths, e.g., the case we considered in the main text, the heat current $J_1$ ($J_2$) flowing into the hot (cold) heat bath is Eq.~(\ref{currentTLSsigmax}) ((\ref{currentTLSsigmaz})). In this case, the populations satisfy a steady-state equation:
\begin{eqnarray}
\left(\Gamma^1_{-(\Omega-\Omega')} + \Gamma^1_{\Omega+\Omega'} + \Gamma^2_{\Omega' }  \right) n_e
=\left(\Gamma^1_{\Omega-\Omega'} + \Gamma^1_{-(\Omega+\Omega')} + \Gamma^2_{-\Omega'}  \right) n_g.
\end{eqnarray}
Carrying out similar discussions for the cases of $\Omega\ge\Omega'$ and $\Omega<\Omega'$, we find that the {\it necessary} condition for a positive steady-state power is
\begin{eqnarray}
\Omega <\left(\frac{\beta_2}{\beta_1}- 1\right) \Omega'.
\end{eqnarray}
If we exchange the coupling modes of these two heat baths and the quantum two-level system, for the case of $\Omega \ge \Omega'$, the quantum machine cannot operate as a heat engine, whereas for the case of $\Omega<\Omega'$, the {\it necessary} condition for the machine operating as a heat engine is
\begin{eqnarray}
\Omega <\left(1-\frac{\beta_1}{\beta_2}\right) \Omega'.
\end{eqnarray}

\section{Eigenvalues of the matrix $\textbf{A}(\vec{\chi})$  and fluctuation theorems}
Substituting the expansion~(\ref{chap6hatrhoA}) into Eq.~(\ref{equationofmotionforrhohat}) and performing simple algebra, we can obtain all elements of the constant matrix $\textbf{A}(\overrightarrow{\chi})$. To clearly see the contributions from the two heat baths, we divide the matrix into two components:
\begin{eqnarray}
\textbf{A}(\overrightarrow{\chi})=\textbf{R}^1(\chi_1^+,\chi_1^-) +\textbf{R}^2(\chi_2^+,\chi_2^-).
\end{eqnarray}
The elements of the first matrix contributed by the hot bath are
\begin{eqnarray}
(\textbf{R}^1)_{11}&=& (e^{i\chi_1^+\hbar\Omega}-1)\Gamma^1_{\Omega}+(e^{-i\chi_1^-\hbar\Omega}-1)\Gamma^1_{-\Omega}-\Gamma^1_{-(\Omega-\Omega')}-\Gamma^1_{\Omega+\Omega'}, \nonumber\\
(\textbf{R}^1)_{12}&=&e^{ i\chi_1^+\hbar(\Omega-\Omega')}\Gamma^1_{\Omega-\Omega'} + e^{- i\chi_1^-\hbar(\Omega+\Omega')}\Gamma^1_{ -(\Omega+\Omega')}, \nonumber \\
(\textbf{R}^1)_{21}&=&e^{-i\chi_1^-\hbar(\Omega-\Omega')}\Gamma^1_{-(\Omega-\Omega')}+ e^{i\chi_1^+\hbar(\Omega+\Omega')}\Gamma^1_{ \Omega+\Omega'},\nonumber \\
(\textbf{R}^1)_{22}&=&(e^{i\chi_1^+\hbar\Omega}-1)\Gamma^1_{ \Omega}  + (e^{-i\chi_1^-\hbar\Omega}-1)\Gamma^1_{ -\Omega} -\Gamma^1_{ \Omega -\Omega'}-
\Gamma^1_{-(\Omega+\Omega')},
\end{eqnarray}
for the case of $\Omega>\Omega'$; otherwise,
\begin{eqnarray}
(\textbf{R}^1)_{12}&=&e^{ i\chi_1^-\hbar(\Omega-\Omega')}\Gamma^1_{\Omega-\Omega'} + e^{- i\chi_1^-\hbar(\Omega+\Omega')}\Gamma^1_{ -(\Omega+\Omega')}, \nonumber \\
(\textbf{R}^1)_{21}&=&e^{-i\chi_1^+\hbar(\Omega-\Omega')}\Gamma^1_{-(\Omega-\Omega')}+ e^{i\chi_1^+\hbar(\Omega+\Omega')}\Gamma^1_{ \Omega+\Omega'},
\end{eqnarray}
whereas $(\textbf{R}^1)_{11}$ and $(\textbf{R}^1)_{22}$ are the same.
The elements of the second matrix contributed by the cold bath are
\begin{eqnarray}
(\textbf{R}^2)_{11}&=& -\Gamma^2_{\Omega'} ,\nonumber\\
(\textbf{R}^2)_{12}&=&  e^{- i\chi_2^-\hbar\Omega' }\Gamma^2_{-\Omega'},\nonumber \\
(\textbf{R}^2)_{21}&=& e^{ i\chi_2^+\hbar\Omega' }\Gamma^2_{\Omega'} , \nonumber \\
(\textbf{R}^2)_{22}&=& -\Gamma^2_{-\Omega'}.
\end{eqnarray}
The eigenvalues of the $2\times 2$ matrix $\textbf{A}(\overrightarrow{\chi})$ are simple:
\begin{eqnarray}
\lambda_\pm(\overrightarrow{\chi})&=&\frac{1}{2}[({\textbf A})_{11}+ ({\textbf A})_{22}\pm B] ,
\end{eqnarray}
where \begin{eqnarray}
B(\overrightarrow{\chi})&=&\sqrt{[({\textbf A})_{11}-({\textbf A})_{22}]^2+4({\textbf A})_{12}({\textbf A})_{21}}.
\end{eqnarray}
Obviously, $\lambda_+>\lambda_-$.

Fluctuation theorems are important for the properties of stochastic efficiencies~\cite{Verley2014a}. According to the concrete expressions of ${\textbf R}^k$ matrixes ($k=1,2$), we can easily verify the following symmetry:
\begin{eqnarray}
\label{matrixAsymmetry}
[\textbf{A}(\chi_1^+,\chi_1^-,\chi_2^+,\chi_2^-)]^T=\textbf{A}(i\beta_1-\chi_1^-,i\beta_1-\chi_1^+,i\beta_2-\chi_2^-,i\beta_2-\chi_2^+),
\end{eqnarray}
where $T$ denotes a transpose. Because the eigenvalues of the transposed matrix are the same as those of the original matrix, we immediately find that the scaled cumulant generation function $\phi(\overrightarrow{\chi})$ satisfies an analogous symmetry:
\begin{eqnarray}
\label{symmetricSCGF}
\phi(\chi_1^+,\chi_1^-,\chi_2^+ ,\chi_2^-)= \phi(-\beta_1-\chi_1^-,-\beta_1-\chi_1^+,-\beta_2-\chi_2^-,-\beta_2-\chi_2^+).
\end{eqnarray}
If we set $\chi_k^+=\chi_k^-=\chi_k$ ($k=1,2$), the above equation is simply the famous Gallavotti-Cohen fluctuation theorem~\cite{Gallavotti1995,Kurchan1998,Lebowitz1999,Maes1999}. Because the original theorem is about the joint probability of the net heat currents while Eq.~(\ref{symmetricSCGF}) is about the joint probability of the absorbed and released heat currents, we call the latter the detailed Gallavotti-Cohen fluctuation theorem~\cite{Liu2020a}. Eq.~(\ref{symmetricSCGF}) explains the steady-state fluctuation theorem of the power $\wp$ and heat current $j_1$~\cite{Verley2014a,Cuetara2015}:
\begin{eqnarray}
\widetilde{\phi}_s(\chi_a,\chi_b)&=& \phi(\chi_a\eta_C-\chi_b,\chi_a\eta_C-\chi_b ,-\chi_b ,-\chi_b)\nonumber\\
&=& \phi(-\beta_1-\chi_a\eta_C+\chi_b,-\beta_1-\chi_a\eta_C+\chi_b ,-\beta_2+\chi_b ,-\beta_2+\chi_b)\nonumber\\
&=& \widetilde{\phi}_s (\beta_2-\chi_a,\beta_2-\chi_b ).
\end{eqnarray}
In addition, we also see why there are no fluctuation theorems for the power $\wp$ and the absorbed total heat current $j^-$:
\begin{eqnarray}
\widetilde{\phi}_d(\chi_a,\chi_b)&=& \phi(-\chi_b, \chi_a\eta_C-\chi_b, -\chi_b, \chi_a\eta_C-\chi_b)\nonumber\\
&=& \phi(-\beta_1- \chi_a\eta_C+\chi_b,-\beta_1-\chi_b, -\beta_2-\chi_a\eta_C+\chi_b,-\beta_2+\chi_b, )\nonumber\\
&=& \widetilde{\phi}_d (\chi_a',\chi_b' ).
\end{eqnarray}
Because $\beta_1\neq \beta_2$, no such $\chi_a'$ and $\chi_b'$ can satisfy this equation.

%\bibliography{RFsubmission20200106}.
%

\end{document}